# Title: Tuning ion coordination preferences to enable selective permeation

**Authors:** Sameer Varma[1] & Susan B. Rempe[1*]
[1]*Computational Bioscience Department, Sandia National Laboratories, Albuquerque, NM-87185*

[*]**Corresponding Author:** Susan B. Rempe, MS 0895, PO Box 5800, Albuquerque, NM-87185, Email: slrempe@sandia.gov, Phone: (505) 845-0253, Fax: (505) 284-3775

**Abbreviations:** Root Mean Square Deviation (RMSD), Glycine dipeptide (GG)

**Acknowledgements:** This work was supported by the Department of Energy. Sandia is a multiprogram laboratory operated by Sandia Corporation, a Lockheed Martin Company, for the U.S. Dept. of Energy. We would like to thank Dr. E Jakobsson for providing us with compute time to carry out simulations on the Intel® Itanium 2 Linux cluster of the National Center for Supercomputing Applications (NCSA) at the University of Illinois, Urbana-Champaign.

**Abstract:** Potassium (K-) channels catalyze $K^+$ ion permeation across cellular membranes while simultaneously discriminating their permeation over $Na^+$ ions by more than a factor of a thousand[1]. Structural studies[2-6] show bare $K^+$ ions occupying the narrowest channel regions in a state of high coordination by all 8 surrounding oxygen ligands from the channel walls. As in most channels, the driving force for selectivity occurs when one ion is preferentially stabilized or destabilized by the channel compared to water[1]. In the common view of mechanism, made vivid by textbook graphics, the driving force for selectivity in K- channels arises by a fit, whereby the channel induces $K^+$ ions to leave water by offering an environment like water for $K^+$, in terms of both energy and local structure. The implication that knowledge of local ion coordination in a liquid environment translates to design parameters in a protein ion channel, producing similar energetic stabilities, has gone unchallenged, presumably due in part to lack of consensus regarding ion coordination structures in liquid water. Growing evidence[7-10] that smaller numbers and different arrangements of ligands coordinate $K^+$ ions in liquid water, however, raises new questions regarding mechanism: how and why should ion coordination preferences change, and how does that alter the current notions of ion selectivity? Our studies lead to a new channel-centric paradigm for the mechanism of $K^+$ ion channel selectivity. Because the channel environment is *not* liquid-like, the channel necessarily induces local structural changes in ion coordination preferences that enable structural and energetic differentiation between ions.

## Manuscript

Several critical processes in living organisms, such as generation of electrical signals or maintenance of osmotic balances and pH, are carried out via movements of ions across cellular membranes. The age-old quest to relate structure to function takes on new significance in these systems because an understanding of mechanism can yield drug targets for a broad spectrum of diseases. From the perspective of engineering, a mechanistic understanding of ion channels can yield design specifications for new classes of materials that mimic exquisitely-controlled biological channel function, thus outperforming current solutions to such diverse problems as water desalination and electrical power generation. Potassium channels provide an exemplary system for study of ion channel mechanism due to the fact that both their structure and function, at least independently, are well defined.

The signature structural element that defines the pore region of a K-channel is its selectivity filter[1,2,11,12]. Figure 1 illustrates a partial view of the selectivity filter of KcsA, where one of the crystallographically resolved $K^+$ ions is highlighted to show its coordination by 8 backbone carbonyl groups arranged in a skewed cubic structure. This key structure of ligands formed by the S1-S3 binding sites to stabilize a $K^+$ ion is understood to mimic the local environment defined in water in terms of both number and geometrical arrangements of coordinating water molecules, and is presumed to generate enough of a structural mismatch to exclude other ions, for instance $Na^+$ ions. There is now growing evidence[7-10], however, that smaller numbers (< 8) and different arrangements of ligands coordinate $K^+$ ions in liquid water, and more specifically, that 8 water molecules seldom coordinate a $K^+$ ion simultaneously in liquid phase (see supplementary text for more discussion). This raises several new questions regarding ion permeation mechanism. Firstly, how and why should ion coordination preferences be different in liquid water and in K-channels, and for that matter, why has evolution selected 8 ligands to coordinate with $K^+$ ions, and not any other "magic" number? Secondly, how do $K^+$ ions permeate the channel at high diffusion rates when the energetic probability that a $K^+$ ion forms 8-fold complexes is negligible? And thirdly, how do these new ideas about structural transitions affect the current notions (for recent reviews see for example Ref. [13-15]) of ion selectivity?

In order to address these issues, we carry out theoretical investigations to interrogate simultaneously both the structural and energetic stabilities of $K^+$ and $Na^+$ ions in liquids and reduced models of K-channels. Consider the following general coordination (cluster) reaction, which also occurs more specifically in the selectivity filter of K-channels:

$$A^+_{gas} + n[X]_M \xrightleftharpoons{\Delta G_{A^+ \rightarrow A^+(X)_n}(M)} [A^+X_n]_M \qquad [1]$$

In this reaction, which takes place in a certain environment $M$, $n$ independent ligands ($X$) from this environment react with monovalent cation $A^+$ to form a coordination complex $\left[A^+X_n\right]_M$. Here, $\left(\Delta G_{A^+ \rightarrow A^+(X)_n}(M)\right)$ is the free energy of this cluster reaction. For example, $\left(\Delta G_{K^+ \rightarrow K^+(H_2O)_3}(L)\right)$ represents the free energy of transferring a $K^+$ ion from gas phase into liquid water when it coordinates with exactly 3 water molecules. For the purpose of understanding selective ion permeation in K-channels, we investigate the thermodynamics of these coordination reactions of $K^+$ and $Na^+$ ions as a function of the number ($n$) and variety ($X$) of the coordinating ligands, concurrently with changes in the environment ($M$) external to the cluster. We calculate the free energies of these reactions using a quasi-chemical approach[16-21], its most significant advantage being that its statistical mechanical construct allows for a quantum treatment of the most important (nearest neighbor) interactions with each ion.

## Solvation free energies in liquid phase

We first consider coordination reactions of $K^+$ and $Na^+$ ions in liquid water. Figure 2 illustrates ion coordination free energies as a function of the number of water molecules in their inner coordination shells. Consistent with other results[7-10], we find that in bulk water, $K^+$ ions do *not* prefer to be associated with 8 oxygen atoms in their inner coordination shells, as otherwise found in the selectivity filters of K-channels. Instead both ions prefer to coordinate with tetrahedral arrangements of 4 water molecules (as illustrated in Figure 1). Also consistent with experimental data[8,10], we find that the water-oxygen atoms are only ~ 0.4 Å further away from the $K^+$ ion as compared to their distances from the $Na^+$ ion, and energetically a hydrated $K^+$ ion is significantly less stable than a hydrated $Na^+$ ion (by ~19 kcal/mol). Note that figure 2 does not indicate the solvation free energies of a $Na^+$ ion associated with more than 6 water molecules, or a $K^+$ with more than 8, in the inner coordination shells. Attempts to optimize higher numbers of water molecules around the ions, such that all the water molecules simultaneously coordinate with the ion, fail. This only implies that the higher coordination states of these ions are energetically less favorable than their respective highest order states noted above (see supplementary text for further discussion).

Since coordination reactions of $K^+$ ions in the selectivity filter of K-channels actually take place with carbonyl oxygen atoms, one wonders if the chemical composition of a carbonyl oxygen atom (in particular its ligand field strength) is different enough from that of a water oxygen atom to induce altered structural and energetic preferences in $K^+$ and $Na^+$ ion coordination. In order to understand the effect of chemical differences between these oxygen atoms, we determine the solvation free energies of both ions in liquid formamide ($NH_2CHO$). The results discussed above in the case of liquid water remain unchanged in liquid formamide. Structures, energy differences, and experimental comparisons match (see supplementary text). We therefore now focus on understanding the effect of the external environment on these coordination reactions with respect to the carbonyl ligands in formamide molecules.

**Solvation free energies in quasi-liquid phase**

In the studies just described, we considered an environment outside the first coordination shell of the ions to be that of a high dielectric medium, which is the natural environment of formamide in its liquid state. We now consider a model situation where we turn off all the electrostatic components of this external environment by setting the dielectric constant of the external environment to unity. In this model situation, we continue to consider a density of formamide molecules equivalent to its density in the liquid state. We do this because a simple calculation shows that the concentration of carbonyl groups in the channel filter (~23 M) is equivalent to the concentration of formamide molecules in its liquid state (~25 M). Furthermore we do not alter the thermal entropic components of these reactions, thus we continue to account for local thermal fluctuations, a property established in earlier work[22] as essential to channel function. For the purpose of reference, we call this a quasi-liquid phase. Figure 3 summarizes the effect of the change of environment on coordination number and reaction free energies of both ions. We find that the free energies of these reactions become progressively favorable as the coordination numbers of the ions increase. Now high coordination characterizes both ions; $K^+$ ion favors 8, and $Na^+$ ion 6, carbonyl ligands. Interestingly, this result remains unchanged if coordination reactions of either ion are carried out in quasi-liquid water (see supplementary text). Thus we find that the electrostatic coupling between the cluster and its external environment, by itself, has a significant impact on the coordination chemistries of the two ions, and turning off this electrostatic coupling leads to a characteristic effect on their favorite coordination numbers. What physics underlies the enormous effect of electrostatic coupling?

This effect can be understood in two alternate but mutually consistent ways, one from the perspective of electrostatic energies, and the other from the perspective of electrical forces. The quasi-liquid state differs from a liquid state because it does not account for the polarization effects of the cluster onto its external environment. Nevertheless, our calculations show that this polarization is only a weak function of the coordination number of the ions, an expected result since the net charge of the cluster is constant (equal to the charge of the central ion). More significantly, the quasi-liquid state also differs from a liquid state in that no electrostatic penalties occur with desolvating (extracting) ligands from their respective liquid media to place them into coordination complexes. According to our calculations, the magnitudes of these electrostatic energies are quite large in liquid media, approximately 8.3 kcal/mol for water and 10.7 kcal/mol for formamide molecules. These electrostatic penalties increase linearly with increasing ion coordination. Therefore, it is this elimination of electrostatic penalties from coordination free energies that results in an upward shift in the favorite coordination numbers of these ions. From the perspective of electrical forces, the effect of the external environment can be understood as follows. In a quasi-liquid state, the ligand molecules that are *not* part of a coordination complex *do not* exert attractive forces on ligands directly coordinating the ions, and therefore cannot pull those ligands away from the ions. There is, in essence, an absence of all such competing forces for ligands in the quasi-liquid state, which again results in an upward shift in the coordination numbers of the two ions.

So far, we have presented two extreme cases of the effect of these electrostatic desolvation penalties. In the lowest dielectric environment, these electrostatic penalties are completely absent, and in condensed phases, these penalties are equivalent to their respective statistical averages in liquid phase. Figure 4 shows how intermediate choices of the dielectric constant of the external environment affect the favorite coordination numbers of these ions in quasi-liquid formamide. We see that even if we have to desolvate the formamide molecules from a medium that has a dielectric constant as low as 1.6, both $K^+$ and $Na^+$ ions would energetically prefer to be associated with the low coordination numbers that characterize the liquid state.

Apart from the unexpected result that a quasi-liquid phase promotes a structural transition in $K^+$ ions to a high 8-fold coordination state, we find that it also results in a shift in energy separation between the ions, due to preferential stabilization (~5 kcal/mol) of $K^+$ ions coordinated with formamide ligands. This suggests that transfer of a $K^+$ ion from liquid water to quasi-liquid formamide would be not only energetically favorable, but also accompanied by an increase in coordination number of $K^+$ ions to 8.

**Quasi-liquid environment of the selectivity filter**

One can expect a similar structural and energetic transition to take place in the selectivity filters of K-channels. Functional groups that could compete with the ion and generate an electrostatic cost for extracting 8 carbonyl oxygens from the filter environment to coordinate permeant $K^+$ ions are none other than proton donors. These proton donors could, in principle, belong to either the side chains of amino acids or to the protein backbone itself. In the context of the K-channels of known 3-d structures (KcsA, MthK, KvAP and KirBac1.1), structural data indicates that, surprisingly, no side-chain amino groups exist at relevant distances from the backbone of the selectivity filter (see supplementary material). Furthermore, in all these channels, for any given backbone carbonyl oxygen of the filter, the nearest backbone proton donor is in fact that oxygen's corresponding backbone nitrogen atom from the adjacent monomer (distance < 4.0 Å). Such interactions between backbone atoms of neighboring monomers have not been reported[2-6,13,14] on conducting states of the filter, however, and would most likely result in pinching the filter shut. The next set of nearest backbone proton donors are at a distance greater than 6.0 Å. They belong to the filter helix and are occupied with maintaining the integrity of its 3-d fold. This implies that if these proton donors were to indeed hydrogen bond with the carbonyl oxygens of the filter backbone, structural transitions of the order of 3.0 Å would be required (considering hydrogen bond distance to be 3.0 Å), which may also involve significant torsional angle changes of the backbone structure. This magnitude of structural transition, however, is much greater than what can be expected solely as a consequence of thermal motions (<1 Å as inferred from the B-factors of their respective x-ray data[2-6]). Nevertheless, the very presence of such proton donors in the vicinity of the selectivity filter warrants a certain degree of rigidity (mean deviation < 3.0 Å) for creating a quasi-liquid environment. As long as the relevant proton donors or carbonyl oxygens of the selectivity filter do not undergo structural transitions greater than 3.0 Å, they would not be able to hydrogen bond with each other, and a quasi-liquid environment could be sustained for partitioning $K^+$ ions into high 8-fold coordinated states.

## 8-fold coordination models of $K^+$ ion

To fully understand and ascertain the realistic applicability of these findings to K-channels, it is imperative to investigate the 3-dimensional structures of the 8-fold coordinated clusters of $K^+$ ions. Figures 5a and 5b illustrate, respectively, the optimized 8-fold coordinated structures of $K^+$ ions with water and formamide molecules. In both cases we find 8 molecules arranged in two separate 4-molecule ring-like structures around the $K^+$ ions. The two 4-molecule rings occupy mutually opposite sides of the $K^+$ ion in positions rotated with respect to each other to form a skewed cubic structure. Figure 5d illustrates the x-ray structure of the S2 binding site of the selectivity filter of KcsA. We see clearly that this skewed cubic geometry of ligands also exists in the x-ray structure, which highlights how optimally the geometry of the selectivity filter of K-channels evolved for $K^+$ ions. These 8-fold model clusters of $K^+$ ions, however, exhibit a characteristic structural feature not present in the structures of K-channels. The coordinating molecules in these two 8-fold clusters form hydrogen bonds to each other. In fact, we find that our 5-, 6-, and 7-fold clusters also show a certain degree of hydrogen bonding between ligands.

To understand the implications of these inter-ligand hydrogen bonds in $K^+$ ion clusters, we use *ab initio* optimization to generate alternate 5-, 6-, 7- and 8-fold $K^+$ ion formamide clusters, ensuring that the formamide molecules avoid hydrogen bond interactions. The alternate non-hydrogen bonded 8-fold cluster of $K^+$ ion is illustrated in Figure 5c. As expected, these non-hydrogen bonded models stabilize the coordinated state of $K^+$ ions less than the corresponding hydrogen bonded clusters (see supplementary material for the reaction free energies calculated using these new optimized structures). In fact, we now find that the free energy of transferring a $K^+$ ion from bulk water to this new non-hydrogen bonded 8-fold formamide cluster in quasi-liquid phase is substantially unfavorable, by +11.7 kcal/mol. Furthermore the quasi-liquid phase no longer induces an upward shift in the favorite coordination number of $K^+$ ions. Clearly, there is more to stabilizing $K^+$ ions in 8-fold coordinated states in the selectivity filter than just eliminating electrostatic penalties associated with extraction of any oxygen ligand from a quasi-liquid filter environment. A more realistic model of the filter ligands should clarify these issues.

## Bidentate character of the selectivity filter

We form a more realistic model of K-channel ligands by optimizing the S2 binding site of the selectivity filter in the presence of a $K^+$ ion (see Figure 5d), which interestingly leads to a root mean square deviation (RMSD) from the crystal structure of only 0.4 Å. Again we calculate the free energy for formation of this complex, noting that this is the same coordination reaction defined in equation 1, except that these ligands are bidentate glycine-dipeptide (GG) molecules. Furthermore, we continue to compute these energies in a quasi-liquid external environment since earlier we identified its sheer necessity and presence in K-channels for the purpose of avoiding extra energetic costs in making ligands available for coordination. The different components of the free energy of this reaction are listed in Table 1. For a fair comparison of entropy between bidentate and monodentate carbonyl ligands, we sum concentration effects (that reduce ligand translational entropy) and thermal entropy to give a total entropy of complex formation.

We find that in quasi-liquid phase the 8-carbonyl GG cluster is more stable than any of the *n*-fold coordinated non-hydrogen bonded formamide complexes. In fact, the free energy of formation of this GG cluster is, within computational errors (see supplementary text), equivalent to the hydration free energy of a $K^+$ ion, and is presumably the reason why $K^+$ ions do not "see" a potential barrier during transfer from a low coordination state in bulk water to an 8-fold coordinated state inside the selectivity filter. We note, however, that an accurate estimation of partition free energy of a $K^+$ ion should also account for its interactions with all atoms in the selectivity filter including other permeating ions. An inspection of the thermodynamic components of the reactions listed in Table 1 reveals the basis for how the selectivity filter of a K-channel stabilizes an 8-fold coordinated state of a $K^+$ ion, without the energetic advantage of inter-ligand hydrogen bonds. While the enthalpic component of the reaction of a $K^+$ ion with 4 GG molecules is less favorable than the enthalpic component of its reaction with 8 formamide molecules forming a ring-like structure, more than half of this loss in enthalpy is compensated for by a lesser loss in total entropy. Thus, the $K^+$ ion in the S2 binding site of the selectivity filter forms an 8-fold coordinated structure with 4 bidentate ligands, an environment like water in terms of energetic stability. The bidentate character of the selectivity filter reduces the net entropy loss previously associated with aggregation of hydrogen-bonded ligands, thereby stabilizing the 8-carbonyl complex preferentially over lower order complexes.

Thus we find that K-channels use their quasi-liquid environments in combination with the bidentate character of their carbonyl ligands to reduce the energetic cost of partitioning a $K^+$ ion from a state of low coordination in bulk water to a state of high 8-fold coordination in their selectivity filters, thereby achieving $K^+$ ion conduction at diffusion-limited rates.

## $Na^+$ ion coordination

These findings raise another critical question regarding $K^+$ ion permeation: how and where do $Na^+$ ions fit into this entire energetic and structural coordination scheme that K-channels utilize for catalyzing $K^+$ ion permeation?

To understand this, we apply the same combination of environmental conditions to $Na^+$ ions that previously stabilized a $K^+$ ion in a state of high 8-fold coordination. Using the original undistorted x-ray structure of the S2 binding site, we freeze it and compute the electronic energy (using quantum calculations) of transferring a $Na^+$ ion relative to a $K^+$ ion into this structure. Assuming thermal contributions to the relative free energies cancel and thus need not be estimated, we find that the unoptimized S2 site of the selectivity filter stabilizes $K^+$ over $Na^+$ ions by 7.6 kcal/mol, consistent with experimental estimates[23]. This suggests that electrostatic interactions, the dominant component of the electronic energy, arising from this particular arrangement of 8 carbonyl oxygens are by themselves sufficient to achieve $K^+/Na^+$ selectivity. Moreover, it appears that thermal motions need not be identical for the two ions nor set to zero (as in a rigid filter structure) for this architecture of electrostatic interactions to give rise to ion selectivity. As computed previously[13,22,24-26] using parameterized force-fields, filter selectivity appears to be tolerant of thermal atomic fluctuations (< 1 Å), fluctuations which are in any case absolutely necessary for ion translocation to take place across the narrowest regions of the filter, which otherwise would obstruct the bare ions.

In contrast to results on the unoptimized filter structure, recent work showed[27] that *ab initio* optimization of the S1 site of the selectivity filter of KcsA containing a $Na^+$ ion results in distortion of its skewed cubic structure to the extent that only 6 of 8 total carbonyl oxygens remain within the first coordination shell of the $Na^+$ ion. In the same context[27], distortion of the S1 site reversed the selectivity, causing preferential stabilization of $Na^+$ over $K^+$ ions, by 2.9 kcal/mol. Carrying out a similar optimization of a $Na^+$ ion in the S2 site, we discovered an alternate structural distortion (see supplementary material for an illustration) that resulted in an even larger preferential $Na^+/K^+$ stabilization of 15.3 kcal/mol. In our distorted structure, which underwent an RMS deviation of 1.8 Å relative to the x-ray structure, only 5 out of the 8 carbonyl oxygen atoms remained within the inner coordination shell of the $Na^+$ ion. Taken together, these results suggest that if the structure of the selectivity filter could distort enough to allow a structural rearrangement of this nature and magnitude, $Na^+$ ions would partition into it and be selectively stabilized over $K^+$ ions. In fact, this may explain the sluggish translocations of $Na^+$ ions via "punch-through" mechanisms across some K-channels at high transmembrane voltages[28]. This may also explain observations of $Na^+$ occupation[14] and $Na^+$ conduction[29] that occur in some K-channels upon the removal of permeant $K^+$ ions. We know, however, that this does not occur readily in the selectivity filters of strongly selective K-channels, especially in the presence of $K^+$ ions since they select $K^+$ over $Na^+$ ions by more than a factor of a 1000. Therefore, we argue that the selectivity filter is *not* particularly flexible as the only way to prevent $Na^+$ ions from readily distorting the structure of the selectivity filter is by providing a certain degree of rigidity (RMSD much less than 1.8 Å) to the backbone of the selectivity filter.

In essence, we find that there is a narrow window of flexibility within which the selectivity filter is able to achieve selective $K^+$ ion permeation: on the one hand, if the filter were frozen solid, $K^+$ ions would not permeate the narrowest channel regions, while on the other hand, if it were entirely flexible and liquid-like, no $K^+/Na^+$ selectivity could occur. Note that this mechanism of selective ion permeation relies on the filter's ability to partition $K^+$ ions into states of high 8-fold coordination, instead of low coordination states characteristic of liquid phases of water and formamide where it was *not* possible to differentiate between $K^+$ and $Na^+$ ions on the basis of local coordination structures. Taken together, it seems as though the coordination preferences of $K^+$ ions are altered and raised in the selectivity filters of strongly selective K-channels for the purpose of structurally differentiating between $K^+$ and $Na^+$ ions and achieving $K^+/Na^+$ selectivity. This adds another dimension to early articulations of the ligand field theory of selectivity[30]. While chemistry (in terms of ligand field strength) undoubtedly still plays a role, we show that highly selective ion partitioning also depends on how well the local architecture matches ion coordination preferences in terms of number and arrangements of surrounding ligands.

It is interesting to note that this mechanism of cation selection is consistent with inferences that were previously drawn from a comprehensive sequence alignment of K-channels[12]. This sequence alignment shows that the residue side chains W68 and T72 in KcsA, which contribute to creating a quasi-liquid environment around the selectivity filter by hydrogen bonding with the side chain of the aromatic residue of its GYG triad, are conserved only in strongly selective K-channels. In weakly selective K-channels, such as HYP or pacemaker channels that exhibit a $K^+$ over $Na^+$ selectivity as low as 3:1, the side chains of these residues are consistently replaced by titratable basic side-chains (competing groups). Given the proximity of these residues to the selectivity filter, it seems plausible that they can help decrease the number of carbonyl oxygen atoms exposed to the permeation pathway by directly hydrogen bonding with them, making the channel more permeant to $Na^+$ ions and/or less permeant to $K^+$ ions.

## Concluding remarks

We show that the coordination preferences of $K^+$ and $Na^+$ are not fixed intrinsic properties of these ions, but instead can be altered by environmental conditions. Nature appears to have specially designed the quasi-liquid architectures of K-channels to alter $K^+$ ion coordination for at least one essential purpose: to select $K^+$ over $Na^+$ ions. The mechanism of ion discrimination is not simply an outcome of architectural constraints and nor is it simply a consequence of the carbonyl-ligand dipole strength. As demonstrated here, it is instead *activated* by a precise combination of local environmental conditions (phase) that lead to alterations in ion coordination preferences, and is then *achieved* via both architectural constraints and combined electrical interactions of the ion and participating ligands. It may therefore be more accurately described as a "phase-activated" induced fit mechanism. Figure 6 summarizes our findings in this respect. In future work it would be interesting to explore how such ion coordination preferences are maintained or changed in other biological systems that catalyze selective ion binding and permeation.

## Methods

The quasi-chemical approach[16-21] for estimating coordination reaction free energies $\Delta G_{A^+ \to A^+(X)_n}(M)$ was quantitatively implemented in two steps. In the first step, the gas-phase thermochemical data, which is required for the equilibria in equation 1, was obtained using *ab initio* calculations. The 3-dimensional structures of ligands ($X$) and their respective coordination complexes with ions $\left(A^+(X)_n\right)$were first optimized at the hybrid B3LYP level of density functional theory. Each optimized structure was then used to determine its individual thermochemistry (via normal mode analysis) at a temperature of 298.15K and pressure of 1 atmosphere. These thermochemistries were then used to obtain gas-phase free energy changes of coordination reactions. The *Gausssian 03* suite of programs[31] was used for both these calculations. Other than the coordination reactions involving GG ($CH_3CONHCH_2CONHCH_3$) molecules, all optimization and subsequent frequency calculations were carried out using the following basis sets: a 6-31G* basis set for $Na^+$ ions, a 6-31++G** basis set for H, C, N & O atoms, and a PNNL basis set for $K^+$ ions. For coordination reactions involving GG molecules, all H, C, N and O atoms were represented using a smaller 6-31G* basis set for computational feasibility. The gas-phase coordination free energies obtained using this procedure were found to be consistent with documented[32,33] experimental values. In the second step, the effect of the environment ($M$) surrounding the complex was computed. The gain in the coordination reaction free energy associated with a higher concentration of ligands in aqueous phase was accounted for implicitly[19-21] as a function of the ratio of their respective concentrations in aqueous and gas phases. The concentration of aqueous water was taken as 55.6 M and that of aqueous formamide was taken as 25.2 M. The concentration of GG was calculated from the concentration of carbonyl groups present along the wall of the selectivity filter of KcsA and was taken as 11.4 M. The electrostatic effect of the surrounding environment was accounted for by treating it as an implicit solvent and numerically solving[34] Poisson's equation using the APBS package[35]. The partial charges of atoms required for creating the charge-grid were obtained using the ChelpG method[36], and the atomic radii required for constructing the solvent exclusion grid were taken from Stefanovich & Truong[37]. The cluster radii needed as per the quasi-chemical rule[18] for partitioning inner shell and outer sphere contributions was taken as the average of the first minima and the second maxima of respective radial distribution plots[9,38,39]. The resulting absolute hydration free energies of both $K^+$ and $Na^+$ ions (Figure 2) were found to be consistent with experimental data (within 2% error).

## Tables

**Table 1:** Different components of the free energy of formation of 8-fold coordination complexes $[K^{+}X_{n}]_{M}$. $\Delta H$ refers to the change in enthalpy; $T\Delta S$ the change in entropy; and $\Delta G(qL)$ refers to the free energy change in the quasi-liquid phase after accounting for the difference between the concentrations of ligands in gas ($C(g)$) and liquid phase ($C(L)$). The concentration of formamide was taken as 0.041 M in gas phase and 25.2 M in liquid phase. The concentration of GG molecules was taken as 0.041M in gas and 11.4 M in condensed phase. All energies are in units of kcal/mol.

| ***Ligands (X)*** | ***n*** | $\Delta H$ | $-T\Delta S$ $(T = 298.15K)$ | $n \cdot \ln\left(C(g)/C(L)\right)$ | $\Delta G(qL)$ |
|---|---|---|---|---|---|
| **Formamide** (H-bonded) | 8 | -127.6 | 81.2 | -31.4 | -77.8 |
| **Formamide** (Not H-bonded) | 8 | -101.8 | 72.5 | -31.4 | -60.7 |
| **GG** | 4 | -107.3 | 50.8 | -13.8 | -70.3 |

## Figure Legends

**Figure 1:** 3-d representations of the coordinated structures of a $K^+$ ion in bulk water and in the selectivity filter of K- channel KcsA. The selectivity filter of KcsA offers 3 binding sites for $K^+$ ions, S1, S2 & S3, where it coordinates with the $K^+$ ion using 8 of its backbone carbonyl oxygen atoms (highlighted in red). Note that the representation of the selectivity filter of KcsA provides only a partial view of its actual 3-d structure. This illustration was created using PyMoL.

**Figure 2:** Reaction free energies of $K^+$ and $Na^+$ ions in liquid water.

**Figure 3:** Reaction free energies of $K^+$ and $Na^+$ ions in quasi-liquid formamide.

**Figure 4:** Effect of the dielectric properties of the external environment on the favorite coordination numbers of $K^+$ and $Na^+$ ions.

**Figure 5:** 3-d representations of the 8-fold coordination structures of $K^+$ ion. The oxygen atoms are highlighted in red, nitrogen atoms in blue, carbon atoms in green and the $K^+$ ions in gold. (a) Optimum structure of 8 water molecules around a $K^+$ ion. The nearest neighbor water molecules are hydrogen bonded (H-bonded) to each other, as indicated by the distance (in Å) between the oxygen atoms. (b) Optimum structure of 8 formamide molecules around a $K^+$ ion. (c) A different optimized structure of a $K^+$ ion coordinating with 8 formamide molecules. This structure differs from the 8-fold formamide cluster illustrated in (b) with respect to the relative distance between the proton donor ($NH_2$ group) of one formamide molecule and the oxygen atom of its nearest neighbor formamide molecule. (d) The x-ray structure of the S2 site of the selectivity filter of KcsA. Refer to text for relevance of these structures to $K^+$ ion permeation. The embedded illustrations were created using PyMoL.

**Figure 6:** Phase-activated induced fit mechanism of ion selectivity. Phase diagrams (a) and (b) illustrate the structural and thermo-chemical effects of transferring $K^+$ and $Na^+$ ions from bulk water into two different quasi-liquid external environments. The absolute free energies of solvating the ions in these external environments can be inferred from the scale on the y-axes of each plot; coordination numbers corresponding to these free energies are indicated in braces. Phase diagram (c) illustrates the thermal effects of transferring these ions from bulk water into quasi-liquid formamide, with the constraints that the ions coordinate with a given number of formamide molecules, which may not form hydrogen bonds to each other. Phase diagram (d) illustrates the effect on the free energy of transferring these ions from bulk water into a quasi-liquid environment where they are coordinating with exactly 4 bidentate GG ligands, where the open symbol denotes the only case where ligands are held rigid. Note that the lines (solid or dashed) connecting the free energy points across phases are of no physical significance, and are drawn merely for the purpose of readability.

# Figures

Figure 1

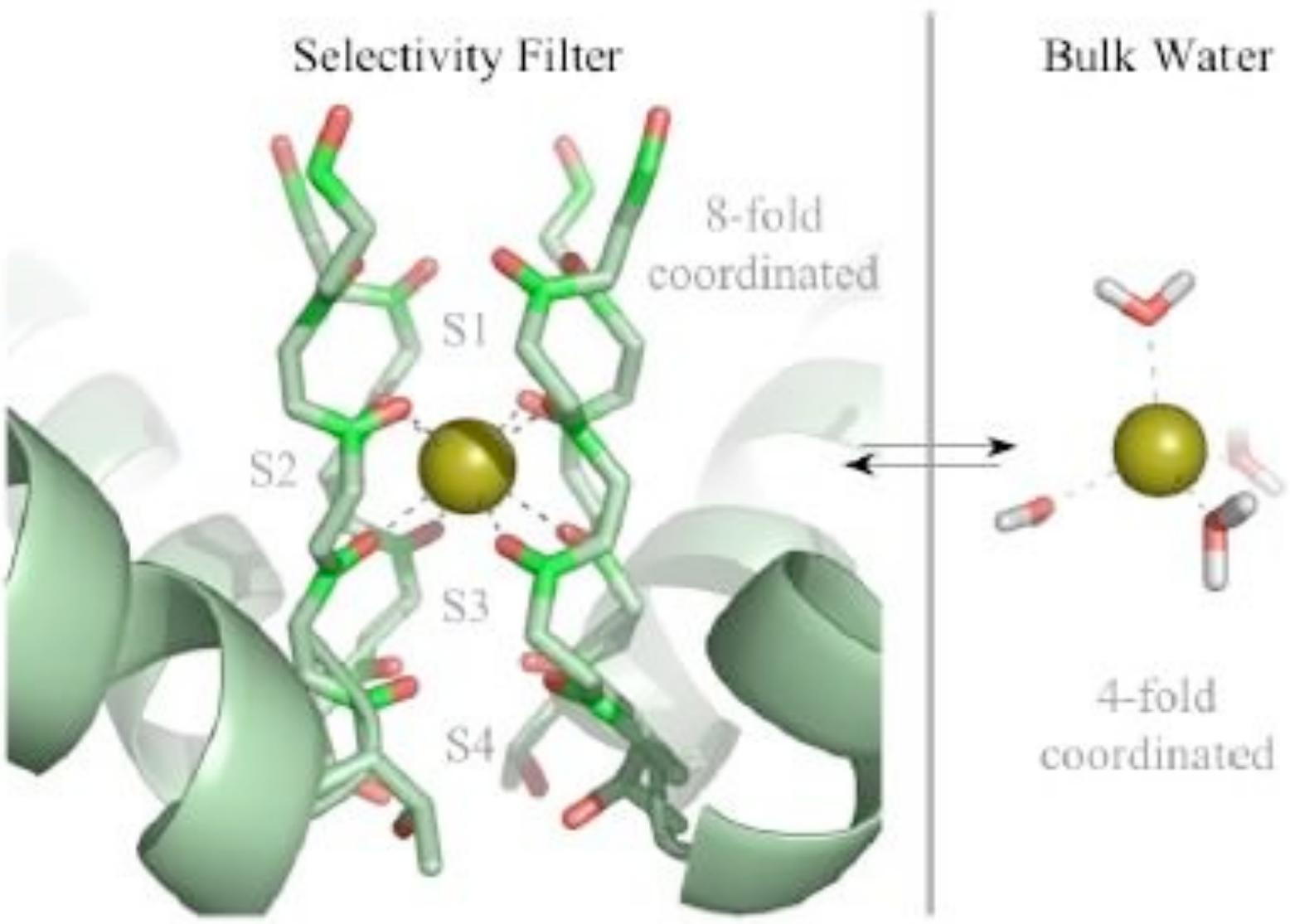

Figure 2

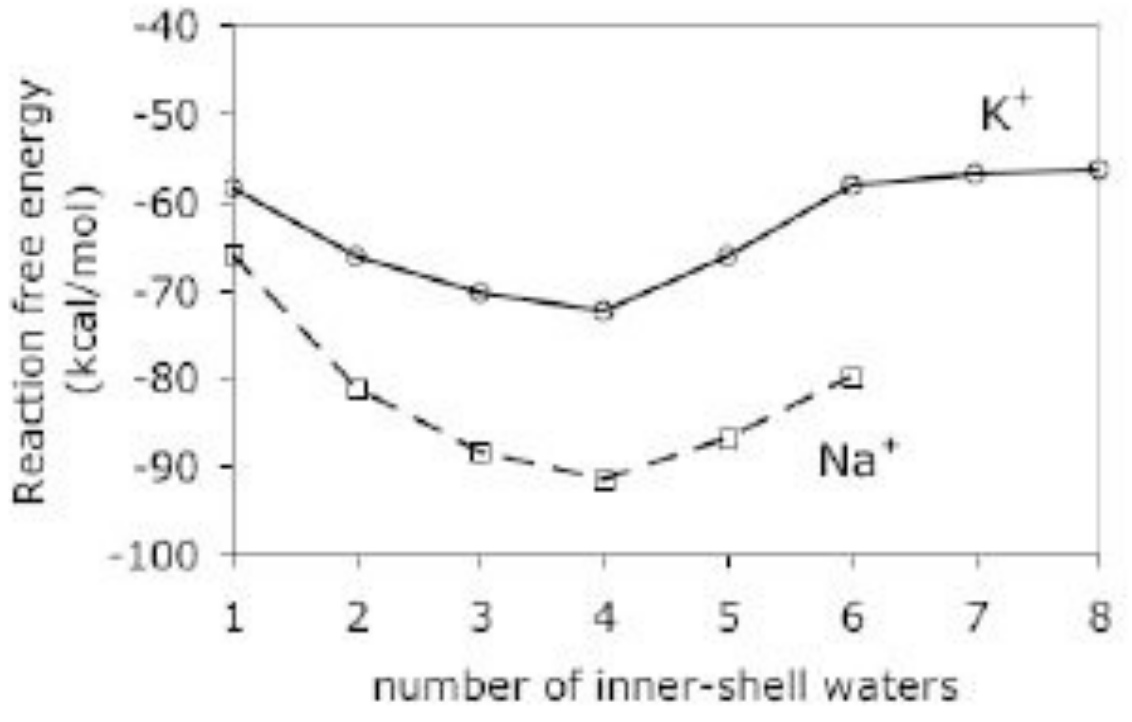

Figure 3

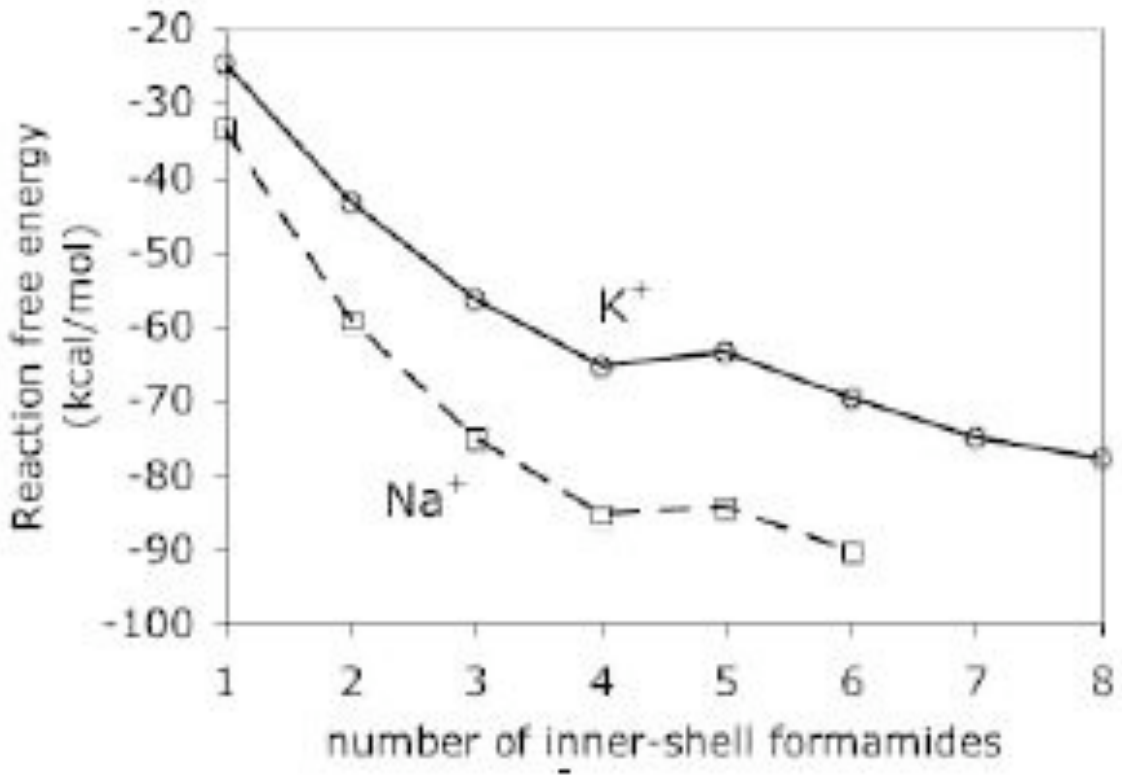

Figure 4

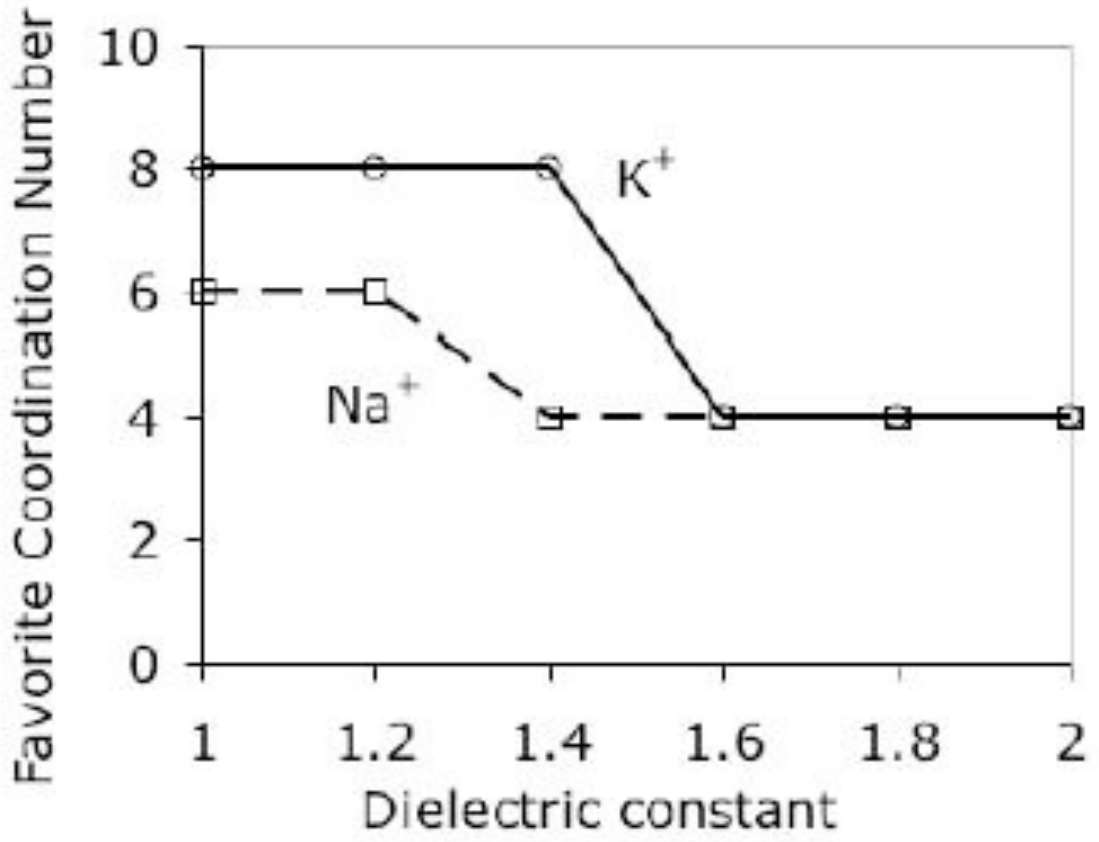

**Figure 5**

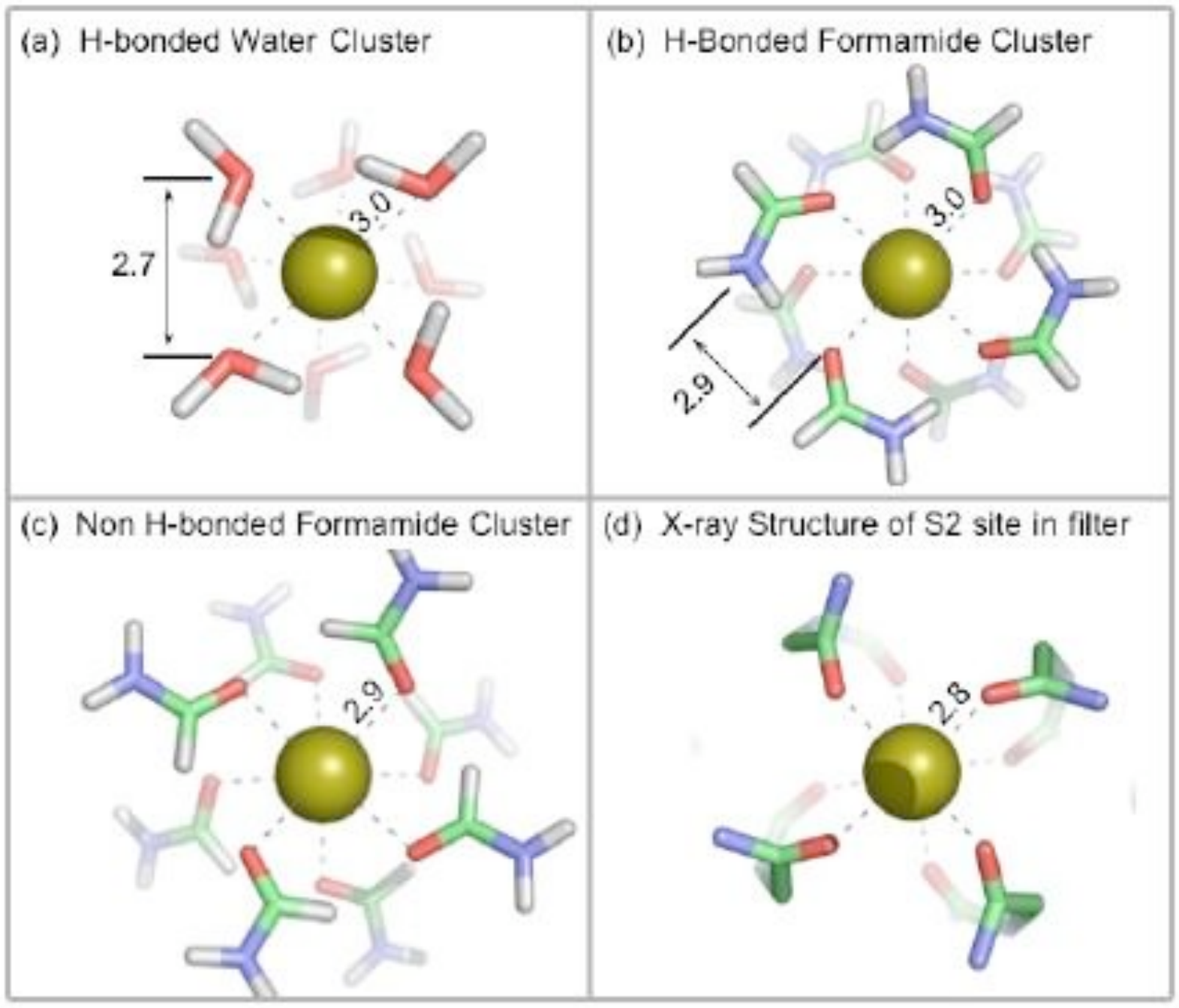

**Figure 6**

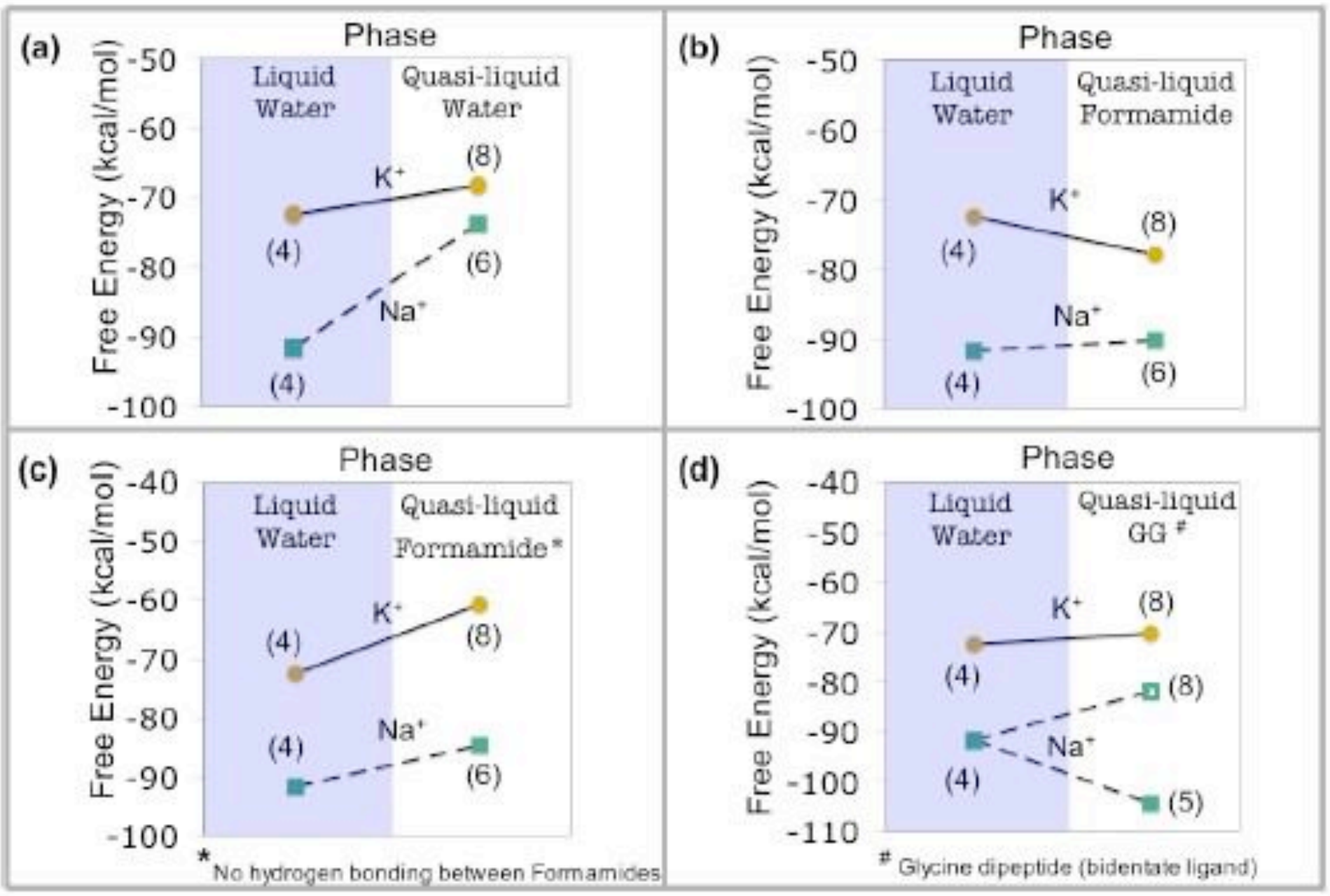

# Supplementary Information

---

**Manuscript Title:** *Tuning ion coordination preferences to enable selective permeation*

**Authors:** *Sameer Varma & Susan B. Rempe*

*Computational Bioscience Department,*

*Sandia National Laboratories, Albuquerque, NM-87158, USA*

**Corresponding Author:** *Susan B. Rempe, MS 0895, PO Box 5800, Albuquerque, NM 87185, Email: slrempe@sandia.gov, Phone: (505) 845-0253, Fax: (505) 284-3775*

---

## There are eight (8) subtopics

### 1) On the most probable coordination structure of ions in water

A lack of consensus characterizes the ion hydration structures determined more than 10 years ago by neutron and x-ray diffraction methods[1,2]. Coordination results reported for $K^+$ and $Na^+$ ions include low (4) and high (8) numbers as well as intermediate possibilities, with neutron diffraction data favoring high coordination. Parameterized force fields for classical molecular simulation generally produce higher ion coordination numbers as well[1,3], although these simulations are not designed to provide the sole determination of such properties.

With improvements in experimental techniques, data analysis strategies, and computational and theoretical approaches for interrogating ion hydration structure, there is now growing evidence[4-13] that smaller numbers and different arrangements of ligands coordinate cations in liquid water. Specifically, current neutron diffraction studies report intermediate 6-fold or lower coordination for $K^+$ ions[8,11] and 5-fold coordination for $Na^+$ ions[12], and x-ray diffraction studies similarly give 5-fold coordination for $Na^+$ ions[8]. Furthermore, data analysis techniques in current development offer a new way to analyze diffraction data that avoids commitment to a specific molecular model[13]. When applied to neutron diffraction data sets for $Li^+$ ion, resulting coordination numbers shift downward to low 4-fold values. The same sequence of intermediate to low values characterizes $K^+$[10], $Na^+$[6,7,9], and $Li^+$[4,5,9] hydration structures reported from *ab initio* methods, which enable molecular dynamics simulations without parameterized classical force field models. More detailed analysis of the *ab initio* simulation data accompanied by probes of structure in a theoretical statistical thermodynamics framework, however, consistently yield low numbers for the subset of innermost ligands (described below) coordinating $Li^+$[4], $Na^+$[7], and $K^+$[10] ions. These predictions agree with results from spectroscopic studies of ions in liquid water, where both $Li^+$ and $Na^+$ ions were attributed with low 4-fold tetrahedral coordination by water molecules[14].

In the traditional definition of hydration number, where the first minimum in the ion-water radial distribution function defines the boundary of the first hydration volume, the intermediate value of 6 determined from *ab initio* molecular dynamics simulation for $K^+$ coordination[10] is consistent with the intermediate values reported by neutron diffraction studies[8,11]. A closer look at the simulation data from multiple investigations[10,15,16], however, indicates sub-structuring in the first peak of the radial distribution of water molecules around a $K^+$ ion. When decomposed for more detailed analysis, overlapping populations of water molecules are found with only 4 water molecules held tightly inside the innermost coordination volume[10], an observation noted in an earlier *ab initio* simulation study as well[15]. Therefore, low 4-fold coordination characterizes the average and the most probable innermost coordination structure of $K^+$ ion in liquid water according to both detailed analysis of the *ab initio* simulation data as well as the theoretical analysis described in the main manuscript. Data from the same simulation[10] also suggests that the probability that a $K^+$ ion coordinates simultaneously with more than 4 water molecules decreases with increasing coordination number, yielding a negligible probability that it forms 8-fold complexes (see Figure 1.1).

## 2) High order coordination complexes of $K^+$ and $Na^+$ ions

Attempts to optimize (using *ab initio* calculations) more than 6 water molecules around a $Na^+$ ion, such that all the water molecules simultaneously coordinated with the ion, failed. On the same note, attempts to optimize more than 8 water molecules around a $K^+$ ion, such that all the water molecules simultaneously coordinated with the ion, also failed. These optimization failures, however, do not in any manner imply that $Na^+$ and $K^+$ ions cannot coordinate with more than 6 and 8 ligands respectively. They essentially only imply that the higher order coordinated states of these ions that could not be optimized are energetically less favorable than their respective highest order states noted above. Furthermore, such higher order coordination states of these ions (>8 for $K^+$ ions and >6 for $Na^+$ ions) have also not been reported in previous aqueous phase AIMD simulations[7,10], which suggest that they are indeed energetically less favorable and that these optimization failures are not likely to be artifacts of gas-phase calculations, as carried out here.

At this point, it is also important to note that we are *not* suggesting that these higher order coordination states cannot be optimized for any other variety of ligands. In fact, it appears that the maximum number of ligands that can potentially be optimized inside an ion's inner coordination shell depends on the field strengths of the ligand dipoles, as exemplified by the S4 site of the selectivity filter of KcsA itself. It was recently shown[17] that the S4 site, which provides a combination of 4 carbonyl dipoles and 4 relatively weaker side-chain hydroxyl dipoles to permeating ions, could be optimized in such a manner such that all these 8 dipoles were inside a $Na^+$ ion's inner coordination shell.

Now the finding that a $K^+$ ion can incorporate a greater number of water or formamide molecules inside its inner coordination shell as compared to a $Na^+$ ion appears to be a consequence of its larger ionic size. A $K^+$ ion has two additional filled electron orbitals as compared to a $Na^+$ ion. This causes the ligand molecules to coordinate with it at distances larger (average of ~ 0.4 Å as inferred from our optimized structures) than what a $Na^+$ ion uses to coordinate with its ligands, which decreases the repulsive forces between the ligands and allows for greater numbers to be incorporated inside its inner coordination shell.

## 3) Solvation free energies of ions in liquid formamide

Figure 3.1 illustrates the solvation free energies of $K^+$ and $Na^+$ ions in liquid formamide. We find that just like in liquid water, both ions prefer to be 4-fold coordinated, with the probability of formation of higher order complexes being lower. We also find that the solvation free energies of the ions are equivalent to their respective hydration free energies. We also find that just like in liquid water, we were unable to optimize more than a certain number of formamide molecules inside their inner coordination shells, which is 8 in the case of $K^+$ ions and 6 in the case of $Na^+$ ions. Taken together, we find that in liquid phase, the differences between the chemistries (in particular their ligand field strengths) of carbonyl and water oxygen atoms are not significant enough to result in any particular structural or energetic differences in ion solvation that can lead us to the mechanistic understandings we seek.

There is, however, one difference between the liquid phase ion-water and liquid phase ion-formamide clusters that we note only for interest. We find that the higher order ($n > 4$) ion-formamide clusters are relatively less probable than their corresponding higher order ion-water clusters. This is because the dielectric constant of liquid formamide (109.5) is higher than the dielectric constant of liquid water (78.5). In the calculation of the solvation energy of these ions in liquid phase, one has to account for the electrostatic penalties associated with the extraction of ligands from their respective liquid media. And because the dielectric constant of liquid formamide is higher than that of liquid water, the desolvation penalty associated with extracting a formamide molecule from liquid formamide (10.7 kcal/mol) is higher than that of the desolvation penalty associated with extracting a water molecule from liquid water (8.3 kcal/mol). These electrostatic penalties increase linearly with increase in the number of ligands coordinating the ion, which explains why the large ion-fomamide clusters are energetically less stable than their corresponding size ion-water clusters in liquid phase.

### 4) Solvation free energies of Ions in Quasi-liquid Water

Figure 4.1 illustrates the reaction free energies of $K^+$ and $Na^+$ ions in quasi-liquid water. As in quasi-liquid formamide, $K^+$ ions prefer to be associated with 8 molecules in their inner coordination shells, while $Na^+$ ions prefer to be associated with 6.

### 5) Quasi-liquid environment of the selectivity filter

The orientation of each potassium channel in Figure 5.1 is as viewed from its respective extra-cellular end. Side chains of residues Arg, Lys, Gln, Asn and His, which can potentially compete with the ion for ligands (competing groups), are highlighted in blue. Side chains of residues Asp and Glu, which cannot compete with the ion for ligands (non-competing groups), are highlighted in red. One can see that there are *absolutely no* competing groups present at relevant distances from the selectivity filters of all channels. In contrast to a polar liquid environment, this clearly provides the selectivity filter with a reduction in energetic cost for making its carbonyl ligands available for coordination with ions. We can also see that all these potassium channels have acidic side chains in the vicinity of the selectivity filter. If these acidic residues were charged (deprotonated) in their crystal configurations, they would electrostatically prevent the carbonyl dipoles from pointing in directions towards their respective permeation pathways, and thereby contribute to the formation of a quasi-liquid environment near the selectivity filter. Each embedded illustration was created using PyMoL.

### 6) Absence of inter-ligand hydrogen bonding destabilizes $K^+$ ion clusters

Two separate sets of n-fold clusters were created using *ab initio* optimization of $K^+$ ions with formamide molecules. During the optimization of one set of clusters, which we refer to as non H-bonded clusters, constraints were applied to prohibit the presence of any inter-ligand hydrogen bonds. As a result, the two optimized sets differed from each other with respect to the arrangements of ligands in the 5-, 6-, 7- and 8-fold clusters. Figure 6.1 shows the reaction free energies calculated using the non H-bonded clusters. Figure 6.2 shows how the absence of inter-ligand hydrogen bonds affects the solvation free energies of the clusters through $\Delta\Delta G$, which is the difference between the reaction free energy of the formation of a non H-bonded cluster and its corresponding sized cluster in the alternate set. We see three distinct effects of the absence of inter-ligand hydrogen bonding. Firstly, in a quasi-liquid phase, a non H-bonded cluster is energetically less stable than its corresponding size cluster that contains inter-ligand hydrogen bonds. Secondly, the free energy of transferring a $K^+$ ion from bulk liquid water to a non H-bonded 8-fold formamide cluster in quasi-liquid phase is unfavorable by +11.7 kcal/mol. Finally, a quasi-liquid phase no longer induces an upward shift in the favorite coordination number of $K^+$ ions, as otherwise seen in the case of the H-bonded clusters.

## 7) Computed hydration free energies of ions compared with experimental values.

The table below compares the hydration free energies of ions computed as part of this work using a quasi-chemical approach to the experimental values.

| | **Computed** (kcal/mol) | **Experimental**[22] (kcal/mol) | **% Error** |
|---|---|---|---|
| **$K^+$** | −72.4 | −74.3 | 2.6 |
| **$Na^+$** | −91.6 | −91.0 | 0.7 |
| **$OH^-$** | −104.3 | −106.5 | 2.2 |
| **KOH** | −176.7 | −180.8 | 2.3 |
| **NaOH** | −195.9 | −197.5 | 0.8 |

## 8) $Na^+$ ion in the S2 site of the selectivity filter of KcsA

An *ab initio* optimization of the S2 site containing a $K^+$ ion resulted in only a small structural change, corresponding to an RMSD of 0.4 Å relative to the x-ray structure. The nature of this structural distortion (figure 8.1b) was such that the average distance between the $K^+$ ion and the carbonyl oxygen atoms increased from 2.8 Å to 3.0 Å. Note that this magnitude of structural deviation is within room temperature RMS fluctuations of these atoms as inferred from the B-factors of the crystal structure of KcsA[23].

An *ab initio* optimization of the S2 site containing a $Na^+$ ion resulted in a much larger structural distortion, corresponding to an RMSD of 1.8 Å relative to the x-ray structure. The nature of this structural distortion (figure 8.1c) was such that 3 out of the 8 carbonyl oxygen atoms moved out of the inner coordination shell of the $Na^+$ ion. The solvation free energy computed using this distorted structure of the S2 site shows that it stabilizes a $Na^+$ ion more than what it is stabilized by bulk water, i.e.,

$$g\left(Na^+\right) = \Delta\Delta G_{Na^+\left(H_2O\right)_4 \to Na^+(GG)_4}\left(L \to qL\right) = -13.2 \text{ kcal/mol.}$$

In addition, the free energy of transferring a $Na^+$ ion from bulk water into this distorted S2 site is more favorable than the free energy of transferring a $K^+$ ion from bulk water into an S2 site optimized for a $K^+$ ion, i.e.,

$$g\left(Na^+\right) - g\left(K^+\right) = -15.3 \text{ kcal/mol.}$$

This implies that if a structural distortion corresponding to a RMS deviation of 1.8 Å were allowed for the S2 site of the selectivity filter, the channel would be more selective toward $Na^+$ ions, rather than being selective toward $K^+$ ions. In other words, this implies that there is a certain degree of rigidity (RMSD < 1.8 Å) that is absolutely necessary for the channel filter to maintain its $K^+/Na^+$ selectivity.

# Figure Legends

**Figure 1.1:** Frequency ($X_n$) of formation of $n$-fold $K^+$-water clusters in liquid phase. Trajectory data used for generating the frequency plot was taken from a previously reported *ab initio* simulation[10] of a $K^+$ ion in liquid water. Frequency calculations were carried out using a cut-off distance of 3.1 Å with respect to the position of the $K^+$ ion.

**Figure 3.1:** Reaction free energies of $K^+$ and $Na^+$ ions in liquid formamide.

**Figure 4.1:** Reaction free energies of $K^+$ and $Na^+$ ions in quasi-liquid water.

**Figure 5.1:** Distribution of polar side chains in the vicinities of the selectivity filters of potassium channels: KcsA[18], MthK[19], KvAP[20] and KirBac1.1[21].

**Figure 6.1:** Reaction free energies of the formation of non H-bonded formamide clusters around $K^+$ ions.

**Figure 6.2:** The energetic consequence of the absence of inter-ligand hydrogen bonds. A non H-bonded cluster is energetically less stable than its corresponding size cluster that contains inter-ligand hydrogen bonds.

**Figure 8.1:** Structural and energetic consequences of *ab initio* optimizations on the S2 site of the selectivity filter of KcsA.

# Figures

## Supplementary Fig. 1.1

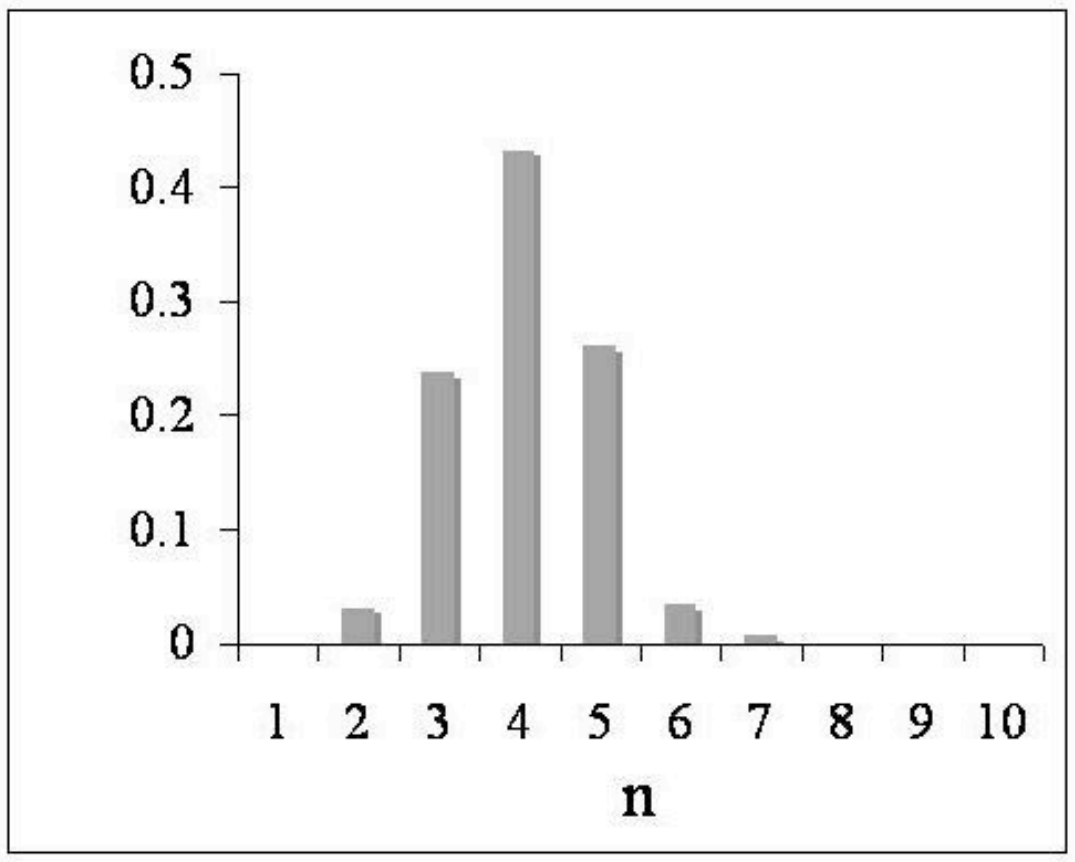

## Supplementary Fig. 3.1

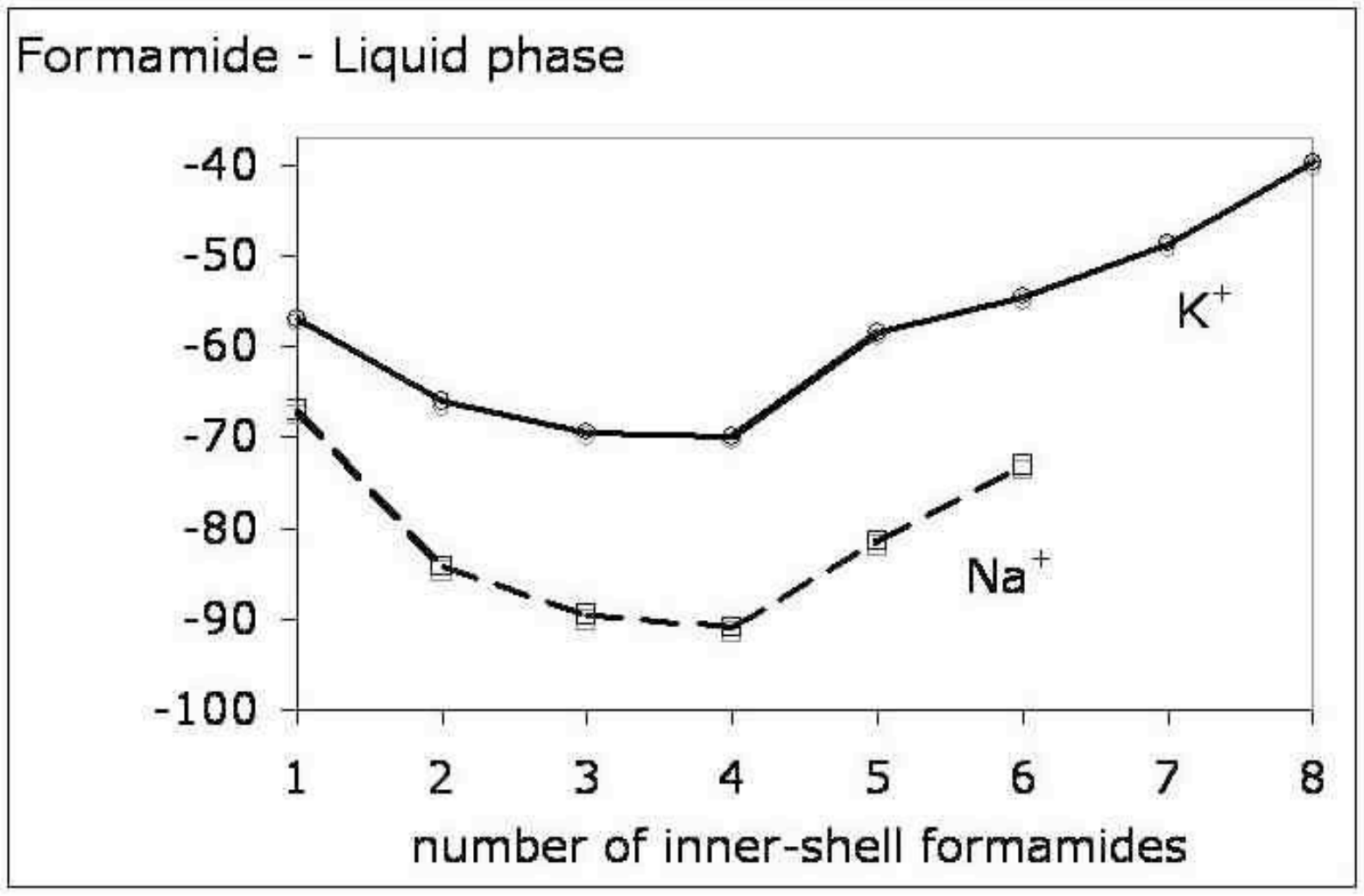

## Supplementary Fig. 4.1

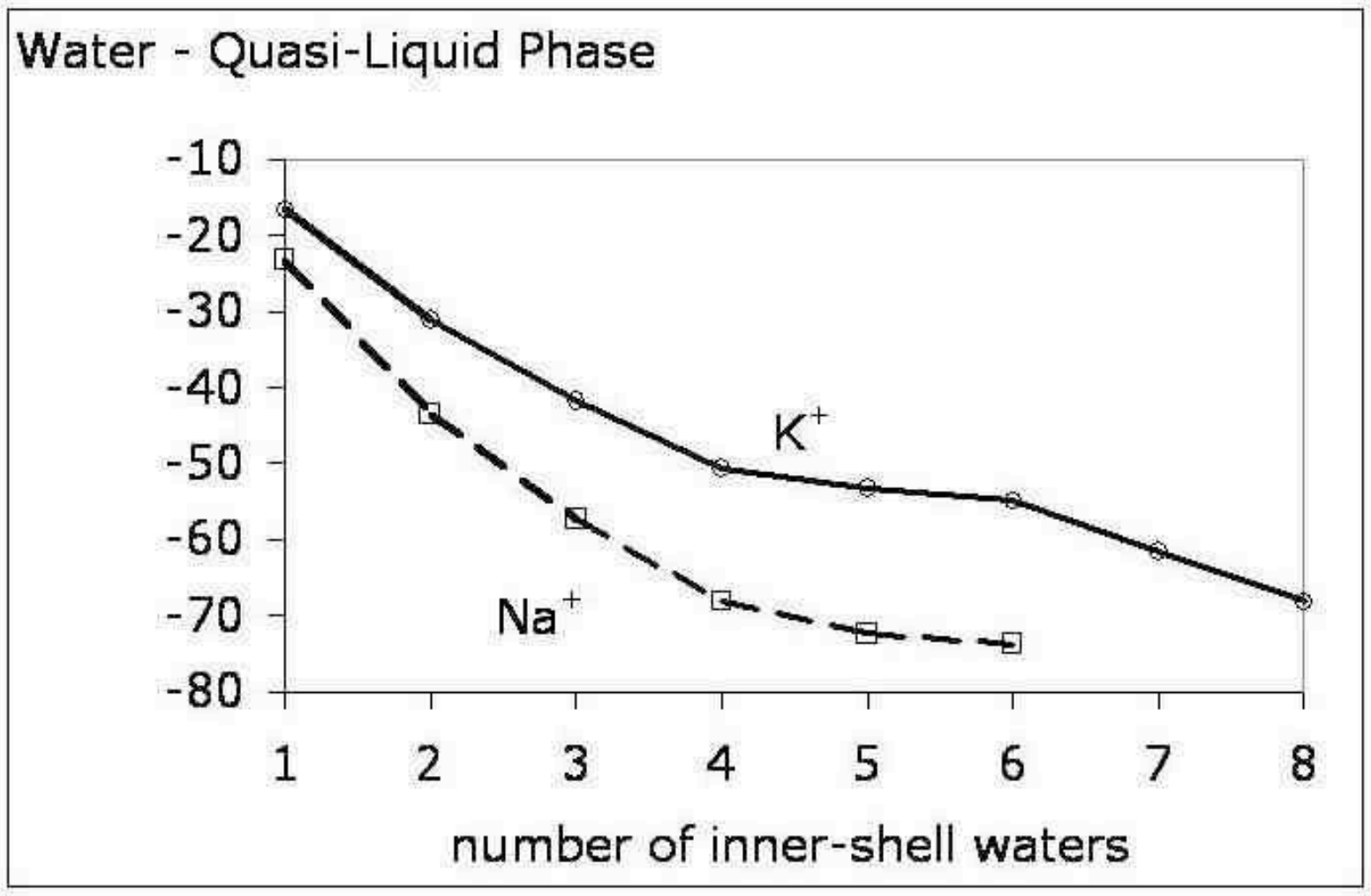

## Supplementary Figure 5.1

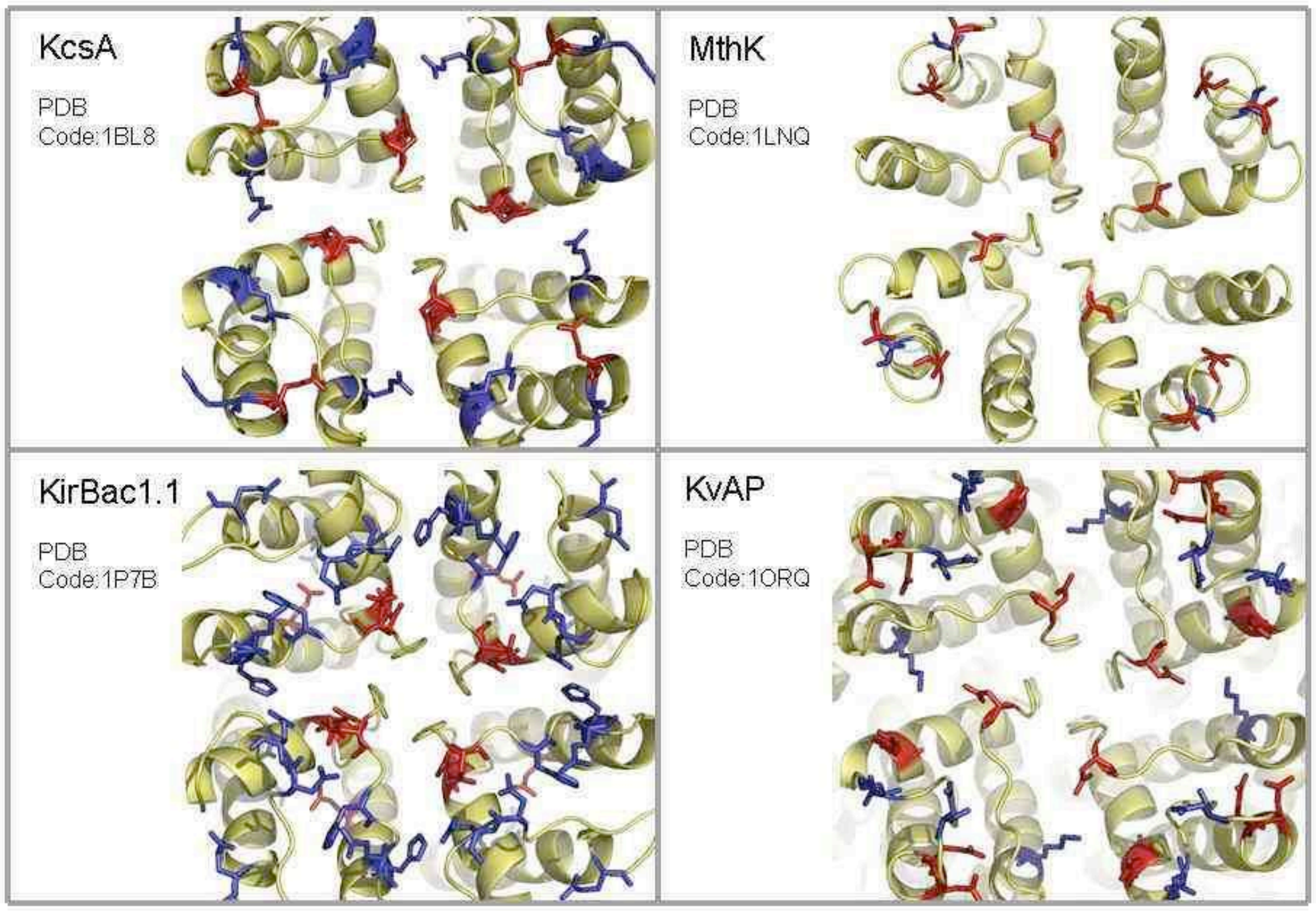

8

**Supplementary Fig. 6.1**

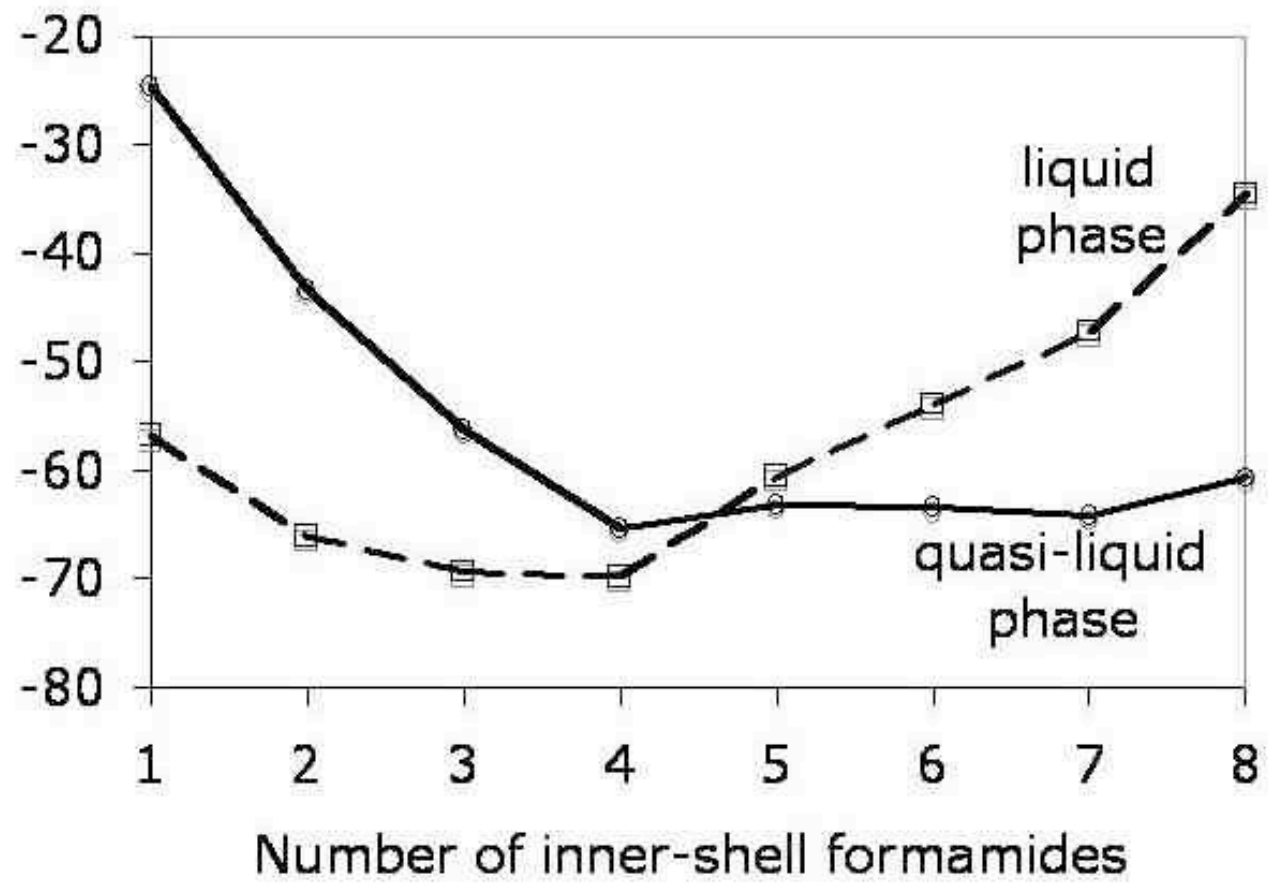

**Supplementary Fig. 6.2**

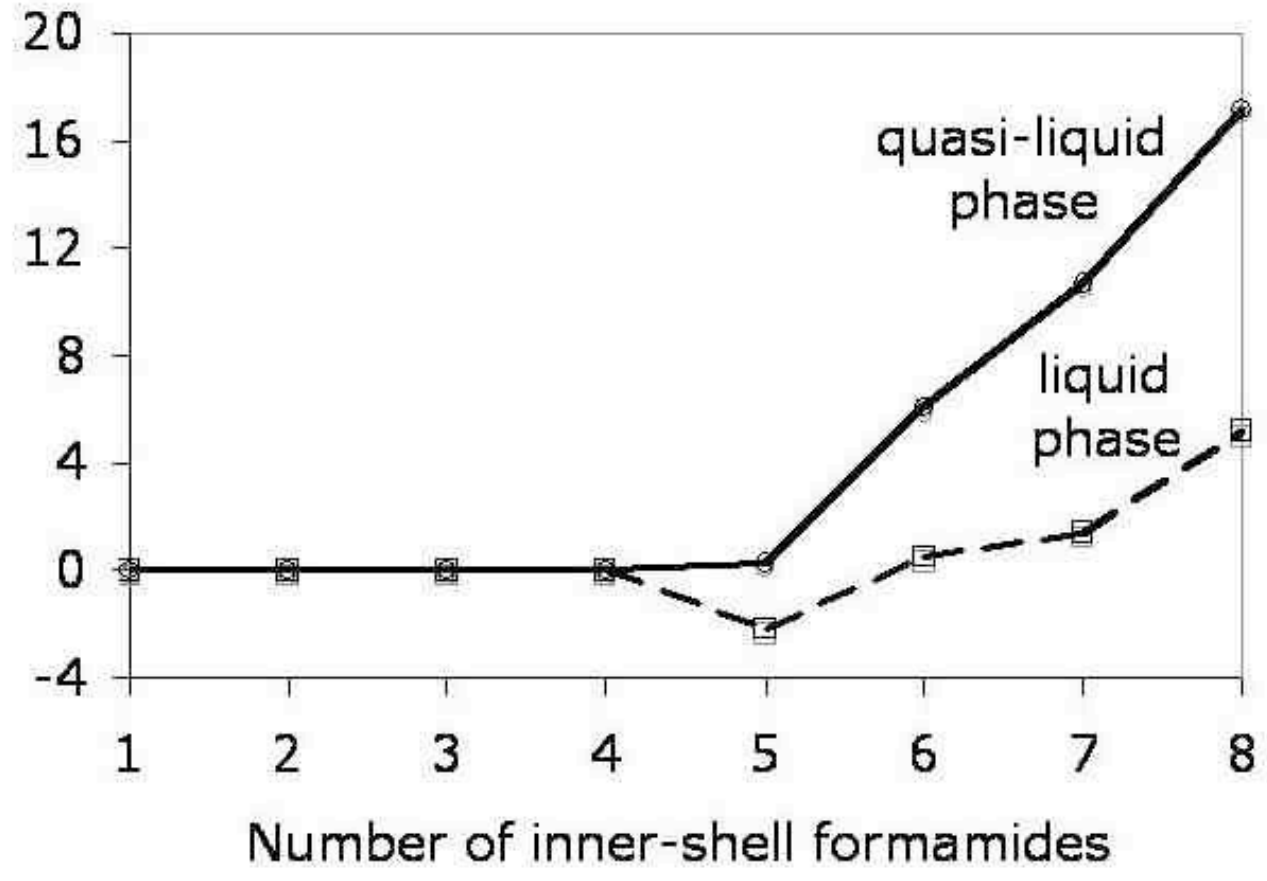

## Supplementary Fig. 8.1

**Ions in S2 site of the selectivity filter of KcsA**

(a) Un-optimized Structure | (b) Optimized with **K⁺** ion | (c) Optimized with **Na⁺** ion

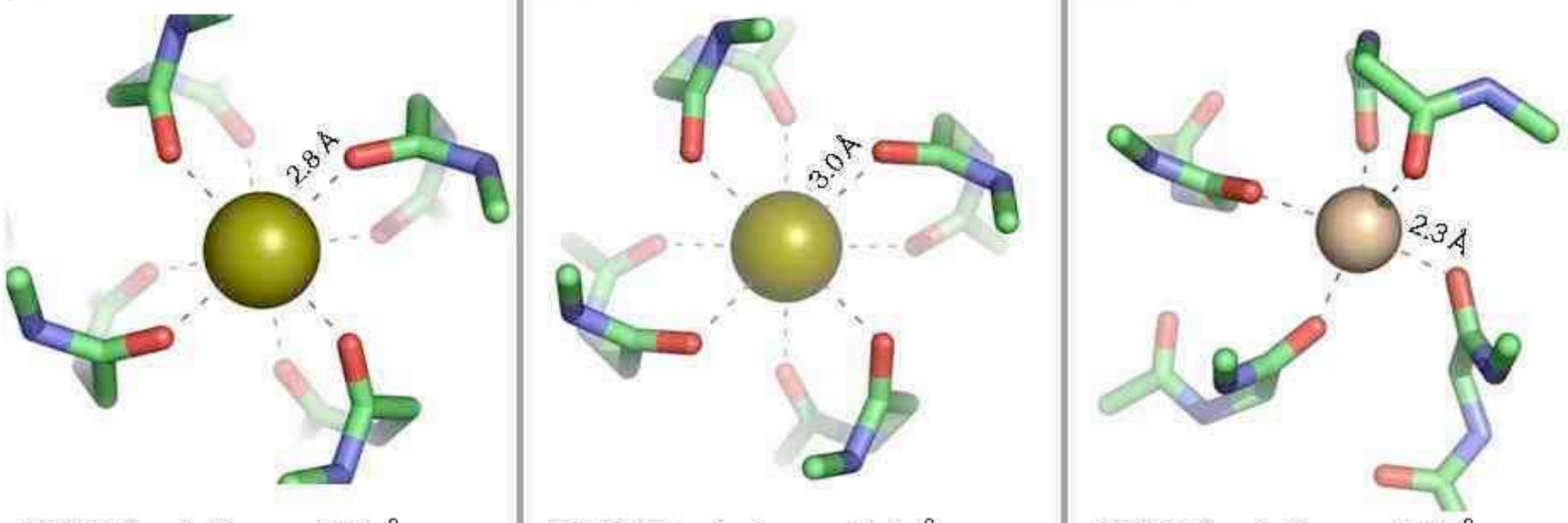

**(a) Un-optimized Structure**

RMS Deviation = 0.0 Å

$$\Delta\Delta G^{electronic}_{K^+(GG)_4 \to Na^+(GG)_4} = \Delta\Delta G^{electronic}_{Na^+(H_2O)_4 \to Na^+(GG)_4} - \Delta\Delta G^{electronic}_{K^+(H_2O)_4 \to K^+(GG)_4}$$

= + 7.6 kcal/mol

**(b) Optimized with $K^+$ ion**

RMS Deviation = 0.4 Å

$$\Delta\Delta G^{electronic}_{K^+(GG)_4 \to Na^+(GG)_4}$$

= + 13.4 kcal/mol

$$\Delta\Delta G_{K^+(H_2O)_4 \to K^+(GG)_4}(L \to qL)$$

= + 2.1 kcal/mol

**(c) Optimized with $Na^+$ ion**

RMS Deviation = 1.8 Å

$$\Delta\Delta G^{electronic}_{K^+(GG)_4 \to Na^+(GG)_4}$$

= - 23.4 kcal/mol

$$\Delta\Delta G_{Na^+(H_2O)_4 \to Na^+(GG)_4}(L \to qL)$$

= - 13.2 kcal/mol